\documentclass[useAMS,usenatbib]{mn2e}
\usepackage{graphicx}
\usepackage{multirow}
\usepackage{dcolumn}
\usepackage{lscape}
\usepackage{array}

\title[]{Surface Thermophysical Properties determination of  OSIRIS-REx target asteroid (101955) Bennu}

\author[LiangLiang Yu \& Jianghui Ji ]{LiangLiang Yu$^{1}$, Jianghui Ji$^{1}$\thanks{jijh@pmo.ac.cn}\\
$^{1}$Key Laboratory of Planetary Sciences, Purple Mountain Observatory, Chinese Academy of Sciences, Nanjing 210008, China}

\begin{document}
\date{Received 2014 December 14; in original form 2014 December 30}

\pagerange{\pageref{firstpage}--\pageref{lastpage}} \pubyear{2002}

\maketitle

\label{firstpage}

\begin{abstract}
In this work, we investigate the thermophysical properties of OSIRIS-REx
target asteroid (101955) Bennu (hereafter, Bennu), where thermal inertia
plays an important role in understanding the nature of the asteroid's surface,
and will definitely provide substantial information for the sampling
return mission. Using a thermophysical model incorporating the
recently updated 3D radar-derived shape model \citep{Nolan2013} and mid-infrared
observations of Spitzer-PUI, Spitzer-IRAC, Herschel/PACS and ESO VLT/VISIR
\citep{Muller2012,Emery2014}, we derive the surface thermophysical properties
of Bennu. The asteroid has an effective diameter of $510^{+6}_{-40}$ m, a geometry
albedo of $0.047^{+0.0083}_{-0.0011}$, a roughness fraction of $0.04^{+0.26}_{-0.04}$,
and thermal inertia of $240^{+440}_{-60}\rm~Jm^{-2}s^{-0.5}K^{-1}$ for a best-fit solution
at 1$\sigma$ level. The best-estimate thermal inertia indicates that fine-grained regolith
may cover a large area of Bennu's surface, with a grain size that may range from
$1.3$ to $31$~mm, and our outcome further supports that Bennu would be a suitable target
for the OSIRIS-REx mission to return samples from the asteroid to Earth.

\end{abstract}


\section{Introduction}
(101955) Bennu (1999 RQ36) is an Apollo-type near-Earth asteroid
(NEA) from a dynamical viewpoint. As is well known,
it is widely believed that Bennu could be a potential Earth impactor
with a relatively high impact probability of approximately $3.7\times10^{-4}$
\citep{Milani2009,Chesley2014}. Nowadays, Bennu is recognized as
one of the potentially hazardous asteroids (PHA) and its orbit
makes it especially accessible for spacecraft. Therefore, NASA
selected Bennu as an ideal target for the OSIRIS-REx sample return
mission \citep{Lauretta2012}, which will be launched in 2016.

On the other hand, based on a linear, featureless spectrum from 0.5
to 2.5 $\mu$m, Bennu is categorized as a B-type asteroid, and the
related spectral analogue is considered to be CM chondrite
meteorites \citep{Bus2002, Clark2011}. B-type asteroids are usually
thought to be primitive and volatile-rich, and most of them are
believed to originate from the middle and outer main belt. Primitive
asteroids can offer key clues for us to understand the process of
planetary formation, the environment of early solar nebulae, and the
emergence of life forms on Earth. From spectroscopic analysis of
B-type asteroids, the surface composition appears to be anhydrous
silicates, hydrated silicates, organic polymers, magnetite, and
sulfides \citep{Larson1983,Clark2010,Ziffer2011,Deleon2012}, whereas
those in the outer main belt may also support the H$_{2}$O-ice stuff
on the subsurface \citep{Rivkin2010,Campins2010}. Both spectral and
dynamical investigations suggest that Bennu may be a liberated
member of the Polana family in the inner main belt
\citep{Campins2010,Bottke2015}.

According to radar observations acquired from Goldstone and Arecibo
at its two closest Earth-approaches in 1999 and 2005, the latest
analysis of \citet{Nolan2013} resolved a nearly spherical shape for
Bennu, where its effective diameter is $492\pm20$ m, the rotation
period is $4.2976\pm0.0002$ h and spin axis is $\beta=-88.0^{\circ}$,
$\lambda=45.0^{\circ}$, indicating a retrograde rotation. The rotation
period is slow enough for the nearly spherical asteroid to allow
potential sampling of the surface regolith. It seems that the rotation
period has not been greatly spun up by tidal or radiation forces
(e.g. YORP effect). However, the recent work of \citet{Chesley2014}
shows that the semimajor axis of Bennu has a mean drift rate
$da/dt=(-19.0\pm0.1)\times10^{-4}~\rm au\cdot Myr^{-1}$ because of
the Yarkovsky effect. On the basis of this estimation, \citet{Chesley2014}
predicted numerous potential impacts for Bennu in the years from
2175 to 2196. It is very necessary to continue to observe
and study the orbital evolution of Bennu. Here we aim to investigate
the thermal inertia of Bennu, because of a close relationship between
thermal inertia and Yarkovsky and YORP effects.

The thermal inertia of an asteroid may be evaluated by fitting
mid-infrared observations with a thermophysical model to reproduce
mid-IR emission curves. \citet{Muller2012} estimated Bennu's thermal
inertia to be $\sim$ $650\rm~Jm^{-2}s^{-0.5}K^{-1}$ with
thermophysical model (TPM), based on observations from
Herschel/PACS, ESO-VISIR, Spitzer-IRS and Spitzer-PUI. Recently,
\citet{Emery2014} showed that a new result of Bennu's thermal
inertia is evaluated to be $310\pm70\rm~Jm^{-2}s^{-0.5}K^{-1}$,
using Spitzer-IRS spectra and a multi-band thermal lightcurve. The
work of \citet{Emery2014} and \citet{Muller2012} differs in two main
aspects: first, the former adopted a 3D radar-derived shape model
\citep{Nolan2013} in their modeling process, whereas the latter
utilized a simple spherical shape model in the investigation;
second, the observational data they used were slightly different, in
that IRAC and IRS peak-up data were included in \citet{Emery2014},
but not utilized in \citet{Muller2012}.

In the present work, we adopt independently developed thermophysical
simulation codes \citep{Yu2014} based on the framework of the Advanced Thermal
Physical Model (ATPM) \citep{Rozitis2011}, to investigate the surface
thermophysical characteristics of Bennu. In our modelling process,
we utilize the radar-derived shape model of asteroid Bennu \citep{Nolan2013}
rather than a spherical approximation shape \citep{Muller2012}. In addition,
the mid-infrared data come from four groups of observations, at various
phase angles, by Spitzer-PUI, Spitzer-IRAC, Herschel/PACS and ESO VLT/VISIR
\citep{Muller2012,Emery2014}. With the rotational averaged ATPM results
fitted to all observations, we provide constraints for thermal inertia
and roughness fraction simultaneously at 1$\sigma$ level. The derived value
of thermal inertia for Bennu is slightly lower than that of \citet{Emery2014},
which may also provide important evidence of the existence of fine-grained
regolith on Bennu's surface. Moreover, on the basis of the derived thermal
inertia, we estimate the grain size of the regolith according to a thermal
conductivity model of \citet{Gundlach2013}. Finally, we summarize the
major outcomes and present a brief discussion.

\section{Thermophysical Modelling}
\subsection{Mid-infrared Observations}
As described above, in our fitting procedure we simply adopt
the observations from Spitzer-PUI, Spitzer-IRAC, Herschel/PACS and
ESO VLT/VISIR \citep{Muller2012,Emery2014}, as there are no
published data sets from Spitzer-IRS. However, the tests of synthetic
data from Spitzer-IRS were generated, and included for fitting to
examine the difference, but we do not find any significant variation in
the results.

If the mid-infrared data at various observational phase angles,
especially at low phase angles, are available, a relatively more
reliable thermal inertia may be derived on the basis of 1D thermal
models. However, currently all published mid-infrared data of Bennu
were observed at large phase angles ($>60^{\circ}$), e.g., those
from Spitzer-IRS, Spitzer-PUI, Spitzer-IRAC, Herschel/PACS and ESO
VLT/VISIR. Although the observations at high phase angles constitute
a disadvantageous condition to constrain thermal inertia and surface
roughness for the asteroid, we can still use these observations, for
a combination fitting to observations performed at several different
phase angles may remove the degeneracy of thermal inertia and
surface roughness in the modelling process, which could somewhat
offset the disadvantage of the lack of low phase angle observations.
From this viewpoint, the absence of Spitzer-IRS observations may not
significantly influence our results, which accord well with the
aforementioned tests with synthetic data. Hence, we tabulate all
data used in the fitting in Table \ref{obs}.

\begin{table*}
\centering
 \renewcommand\arraystretch{1}
\caption{Observational data used in this work (from
\citet{Muller2012} and \citet{Emery2014})} \label{obs}
\begin{tabular}{@{}ccccccccc@{}}
\hline
 UT  & \multicolumn{4}{c}{Flux (mJy)} & $r_{\rm helio}$ & $\Delta_{\rm obs}$ & $\alpha$ & Observatory\ \\
     & \multicolumn{2}{c}{16.0 ($\mu m$)} & \multicolumn{2}{c}{22.0 ($\mu m$)}
     & (AU) & (AU) & ($^{\circ}$) & Instrument \\
\hline
2007-05-03 00:00&\multicolumn{2}{c}{12.49$\pm$0.28}&\multicolumn{2}{c}{11.85$\pm$0.41}&1.12380&0.50564&-63.52&Spitzer-PUI\\
2007-05-03 02:32&\multicolumn{2}{c}{12.70$\pm$0.28}&\multicolumn{2}{c}{11.94$\pm$0.42}&1.12444&0.50650&-63.48&Spitzer-PUI\\
2007-05-03 03:29&\multicolumn{2}{c}{13.38$\pm$0.28}&\multicolumn{2}{c}{12.40$\pm$0.42}&1.12864&0.51222&-63.47&Spitzer-PUI\\
2007-05-03 04:40&\multicolumn{2}{c}{12.69$\pm$0.29}&\multicolumn{2}{c}{12.09$\pm$0.40}&1.12516&0.50748&-63.46&Spitzer-PUI\\
2007-05-03 08:10&\multicolumn{2}{c}{12.87$\pm$0.27}&\multicolumn{2}{c}{12.05$\pm$0.41}&1.12816&0.51155&-63.41&Spitzer-PUI\\
2007-05-03 08:33&\multicolumn{2}{c}{12.47$\pm$0.28}&\multicolumn{2}{c}{11.83$\pm$0.42}&1.12528&0.50762&-63.40&Spitzer-PUI\\
2007-05-03 09:55&\multicolumn{2}{c}{13.00$\pm$0.28}&\multicolumn{2}{c}{12.23$\pm$0.40}&1.12414&0.50611&-63.38&Spitzer-PUI\\
2007-05-03 10:44&\multicolumn{2}{c}{13.00$\pm$0.28}&\multicolumn{2}{c}{12.28$\pm$0.43}&1.12657&0.50938&-63.25&Spitzer-PUI\\
2007-05-03 20:08&\multicolumn{2}{c}{13.08$\pm$0.30}&\multicolumn{2}{c}{12.18$\pm$0.44}&1.12428&0.50629&-63.25&Spitzer-PUI\\
2007-05-04 07:40&\multicolumn{2}{c}{12.85$\pm$0.29}&\multicolumn{2}{c}{12.17$\pm$0.42}&1.12492&0.50716&-63.10&Spitzer-PUI\\
2007-05-04 11:11&\multicolumn{2}{c}{12.97$\pm$0.28}&\multicolumn{2}{c}{12.62$\pm$0.42}&1.12497&0.50722&-63.05&Spitzer-PUI\\
\hline
\\
\\
\hline
 UT & \multicolumn{4}{c}{Flux (mJy)} & $r_{\rm helio}$ & $\Delta_{\rm obs}$ & $\alpha$ & Observatory\ \\
    & \multicolumn{1}{c}{3.6 ($\mu m$)} & \multicolumn{1}{c}{4.5 ($\mu m$)}
    & \multicolumn{1}{c}{5.8 ($\mu m$)} & \multicolumn{1}{c}{8.0 ($\mu m$)}
    & (AU) & (AU) & ($^{\circ}$) & Instrument \\
\hline
2007-05-08 16:24&0.0545$\pm$0.0033&0.324$\pm$0.009&1.396$\pm$0.043&4.968$\pm$0.044&1.14245&0.53224&-61.78 & Spitzer-IRAC\\
2007-05-08 16:46&0.0482$\pm$0.0042&0.327$\pm$0.012&1.396$\pm$0.043&4.983$\pm$0.067&1.14250&0.53232&-61.78 & Spitzer-IRAC\\
2007-05-08 17:14&0.0481$\pm$0.0029&0.332$\pm$0.006&1.495$\pm$0.035&5.063$\pm$0.074&1.14256&0.53241&-61.77 & Spitzer-IRAC\\
2007-05-08 17:39&0.0488$\pm$0.0032&0.334$\pm$0.008&1.397$\pm$0.030&4.970$\pm$0.060&1.14261&0.53250&-61.77 & Spitzer-IRAC\\
2007-05-08 18:06&0.0494$\pm$0.0032&0.331$\pm$0.009&1.382$\pm$0.028&4.876$\pm$0.066&1.14268&0.53259&-61.76 & Spitzer-IRAC\\
2007-05-08 18:27&0.0450$\pm$0.0027&0.335$\pm$0.013&1.340$\pm$0.042&4.985$\pm$0.047&1.14272&0.53266&-61.76 & Spitzer-IRAC\\
2007-05-08 18:59&0.0425$\pm$0.0022&0.308$\pm$0.009&1.353$\pm$0.049&4.664$\pm$0.049&1.14280&0.53278&-61.75 & Spitzer-IRAC\\
2007-05-08 19:22&0.0439$\pm$0.0031&0.330$\pm$0.010&1.399$\pm$0.042&5.048$\pm$0.053&1.14285&0.53285&-61.75 & Spitzer-IRAC\\
2007-05-08 19:50&0.0492$\pm$0.0041&0.356$\pm$0.009&1.540$\pm$0.042&5.203$\pm$0.045&1.14291&0.53295&-61.74 & Spitzer-IRAC\\
2007-05-08 20:16&0.0486$\pm$0.0042&0.317$\pm$0.009&1.370$\pm$0.044&4.907$\pm$0.052&1.14297&0.53304&-61.73 & Spitzer-IRAC\\
2007-05-08 20:39&0.0447$\pm$0.0031&0.316$\pm$0.009&1.435$\pm$0.044&5.018$\pm$0.038&1.14302&0.53312&-61.73 & Spitzer-IRAC\\
\hline
\\
\\
\hline
 UT & \multicolumn{2}{c}{Wavelength} & \multicolumn{2}{c}{Flux }
    & $r_{\rm helio}$ & $\Delta_{\rm obs}$ & $\alpha$ & Observatory\ \\
    & \multicolumn{2}{c}{($\mu m$)}  & \multicolumn{2}{c}{(mJy)}
    & (AU) & (AU) & ($^{\circ}$) & Instrument \\
\hline
2011-09-09 19:01&\multicolumn{2}{c}{ 70.0}&\multicolumn{2}{c}{27.0$\pm$1.7}&1.0146021&0.1742221&+85.3&Herschel/PACS \\
2011-09-09 19:59&\multicolumn{2}{c}{ 70.0}&\multicolumn{2}{c}{24.3$\pm$1.5}&1.0144716&0.1742140&+85.4&Herschel/PACS \\
2011-09-09 19:25&\multicolumn{2}{c}{100.0}&\multicolumn{2}{c}{14.2$\pm$1.1}&1.0145368&0.1742180&+85.4&Herschel/PACS \\
2011-09-09 20:24&\multicolumn{2}{c}{100.0}&\multicolumn{2}{c}{12.0$\pm$1.2}&1.0144064&0.1742101&+85.4&Herschel/PACS \\
2011-09-09 19:25&\multicolumn{2}{c}{160.0}&\multicolumn{2}{c}{ 6.5$\pm$2.3}&1.0145368&0.1742180&+85.4&Herschel/PACS \\
2011-09-09 19:59&\multicolumn{2}{c}{160.0}&\multicolumn{2}{c}{ 4.4$\pm$1.9}&1.0144716&0.1742140&+85.4&Herschel/PACS \\
2011-09-09 20:24&\multicolumn{2}{c}{160.0}&\multicolumn{2}{c}{ 7.6$\pm$2.0}&1.0144064&0.1742101&+85.4&Herschel/PACS \\
2011-09-17 09:21&\multicolumn{2}{c}{ 8.59}&\multicolumn{2}{c}{23.7$\pm$3.6}&0.9908803&0.1798955&+89.4&ESO VLT/VISIR \\
\hline
\end{tabular}
\end{table*}

\subsection{Advanced thermophysical model}
In ATPM \citep{Rozitis2011, Yu2014}, an asteroid is considered to be a
polyhedron composed of $N$ triangle facets. For each facet, the
conservation of energy leads to an instant thermal equilibrium between
sunlight, thermal emission, thermal diffusion, multiple-scattered
sunlight and thermal-radiated fluxes from other facets. If each facet
is small enough, the thermal diffusion on the asteroid can be
approximatively described as one-dimensional (1D) heat diffusion.
Hence, the temperature $T_{i}$ of each facet varies with time as
the asteroid rotates. In this process, $T_{i}$ can be enlarged by
multiple-scattered sunlight and thermal-radiated fluxes from other
facets, which explains well the so-called thermal infrared
beaming effect. When the entire asteroid comes into the final
thermal equilibrium state, $T_{i}$ will change periodically following
the rotation of the asteroid. Therefore, we can build numerical
codes to simulate $T_{i}$ at any rotation phase for the
asteroid. For a given observation epoch, ATPM can reproduce a
theoretical profile to each observation flux as:
\begin{equation}
F_{\rm model}(\lambda)=\sum^{N}_{i=1}\varepsilon f(i)B(\lambda, T_{i})~,
\end{equation}
where $\varepsilon$ is the emissivity, $f(i)$ is the view factor of facet $i$ to the telescope and
$B(\lambda, T_{i})$ is the Plank function:
\begin{equation}
B(\lambda, T_{i})=\frac{2\pi hc^{2}}{\lambda^{5}}\frac{1}{\exp\big(\frac{hc}{\lambda kT_{i}}-1\big)}~.
\end{equation}
Thus the calculated $F_{\rm model}$ can be compared with the thermal
infrared fluxes summarized in Table \ref{obs} in the fitting
process.

\subsection{Fitting Procedure}
In order to derive the thermophysical nature of Bennu via the ATPM
procedure, we need to know several physical parameters - the shape
model, surface roughness, geometric albedo, thermal inertia, thermal
conductivity and thermal emissivity - which are used as the initial
parameters in the calculations. In addition, the heliocentric distance,
geocentric distance and phase angle of Bennu are well determined
because the asteroid's orbit has been accurately measured by optical
observations. On the other hand, we employ the radar-resolved shape
model \citep{Nolan2013} for Bennu in our fitting procedure, and the
incorporation of that shape model will definitely
provide an improved determination of the thermal inertia
\citep{Emery2014,Yu2014}.

According to \citet{Fowler1992}, an asteroid's effective diameter
$D_{\rm eff}$, which is defined by the diameter of a sphere with a
volume identical to what the radar-derived shape model describes,
can be related to its geometric albedo $p_{v}$ through its absolute
visual magnitude $H_{v}$ by the following equation:
\begin{equation}
D_{\rm eff}=\frac{1329\times 10^{-H_{v}/5}}{\sqrt{p_{v}}}~(\rm km) ~.
\label{Deff}
\end{equation}
Thus, we actually have three free parameters - thermal inertia,
roughness fraction and effective diameter (or geometric albedo)-
that should be extensively investigated in the fitting process.
Other parameters are listed in Table \ref{phpa}.

Herein the surface roughness is modeled by a fractional coverage of
hemispherical craters, symbolized by $f_{\rm R}$, whereas the
remaining fraction, $1-f_{\rm R}$, represents a smooth flat surface
on the asteroid. The hemispherical crater adopted in this work is
a low-resolution model consisting of 132 facets and 73 vertices,
following a treatment similar to that shown in \citep{Rozitis2011,Wolters2011,Yu2014}.
As the sunlight is more easily scattered over a rough surface
than a smooth flat region, roughness can decrease the effective Bond
albedo. Using the above mentioned model of surface roughness, the effective
Bond albedo $A_{\rm eff}$ of a rough surface can be relevant to the Bond
albedo $A_{\rm B}$ of a smooth flat surface and the roughness fraction
$f_{\rm R}$ by \citep{Wolters2011,Yu2014}

\begin{equation}
A_{\rm eff}=f_{\rm R}\frac{A_{\rm B}}{2-A_{\rm B}}+(1-f_{\rm R})A_{\rm B}~.
\label{aeffab}
\end{equation}
On the other hand, the effective Bond albedo $A_{\rm eff}$ is related to
geometric albedo $p_{v}$ by
\begin{equation}
A_{\rm eff}=p_{v}q_{\rm ph}~,
\label{aeffpv}
\end{equation}
where $q_{\rm ph}$ is a phase integral that can be approximated by
\citep{Bowell}
\begin{equation}
q_{\rm ph}=0.290+0.684G~,
\label{qph}
\end{equation}
where $G$ is the slope parameter in the $H, G$ magnitude system of
\citet{Bowell}.
Then for each thermal inertia $\Gamma$, roughness fraction $f_{\rm R}$,
and effective diameter $D_{\rm eff}$ case, a flux correction factor $FCF$
is defined as \citep{Wolters2011}
\begin{equation}
FCF=\frac{1-A_{B,now}}{1-A_{B,initial}}~,
\end{equation}
where $A_{B,now}$ is calculated by inversion of equation
(\ref{aeffab}), to fit the observations, and then the so-called
reduced $\chi^{2}$ defined as \citep{Muller2011}
\begin{equation}
\chi^{2}_{\rm reduced}=\frac{1}{n}\sum^{n}_{i=1}\Big(\frac{FCF\cdot F_{\rm model}
    -F_{\rm obs}(\lambda_{i})}{\sigma_{\lambda_{i}}}\Big)^{2}~,
\label{l2}
\end{equation}
can be obtained to assess the fitting degree of our model with
respect to the observations. Herein the predicted model flux $F_{\rm
model}(\Gamma,f_{\rm R},D_{\rm eff},\lambda)$ is a rotationally
averaged profile, since the rotation phase of Bennu was unknown at
the time of observation. In addition, the FCF plays a less
significant role in determining thermal inertia, but simply brings
about $\sim$ 0.2\% influence on the outcome of effective diameter
and geometric albedo.

\begin{table}
 \centering
 \renewcommand\arraystretch{1.2}
 \caption{Assumed physical parameters used in ATPM.}
 \label{phpa}
 \begin{tabular}{@{}lcc@{}}
 \hline
 Property & Value & References \\
 \hline
 Number of vertices  &     1348               & \citep{Nolan2013}  \\
 Number of facets    &     2692               & \citep{Nolan2013}  \\
 Shape (a:b:c)       & 1.1135:1.0534:1        & \citep{Nolan2013}  \\
 Spin axis           & ($-88.0^{\circ}$,$45.0^\circ$) & \citep{Nolan2013} \\
 Spin period         &     4.2976 h           & \citep{Nolan2013}  \\
 Absolute magnitude  &     20.40              & \citep{Hergenrother2013} \\
 Slope parameter     &     -0.08              & \citep{Hergenrother2013} \\
 Emissivity          &      0.9               & assumption \\
 \hline
\end{tabular}
\end{table}

In order to simplify the best-fit searching process, a set of
thermal inertia are given in the range
$0\sim1000\rm~Jm^{-2}s^{-0.5}K^{-1}$. For each case, a series of
roughness fractions and effective diameters are evaluated to find out
which can be taken as the likely solution with respect to the
observations. The fitting outcomes are summarized in Table \ref{fitchi2}.
In this table, the $\chi^{2}$ values are thereby relevant to each thermal
inertia, roughness fraction and effective diameter. Each effective diameter
listed in Table \ref{fitchi2} is a profile that gives a smallest $\chi^{2}$
for each thermal inertia and roughness fraction pair. Roughly speaking,
the $\chi^{2}$ values imply a best-estimate solution of thermal inertia between
$200\sim300\rm~Jm^{-2}s^{-0.5}K^{-1}$, which can be derived from the
smallest $\chi^{2}$ value in the two-dimension phase space of
$\chi^{2}$ ($\Gamma$, $f_{\rm R}$).

\section{Results Analysis}
\subsection{Thermal inertia and Roughness fraction}
\begin{figure}
\includegraphics[scale=0.6]{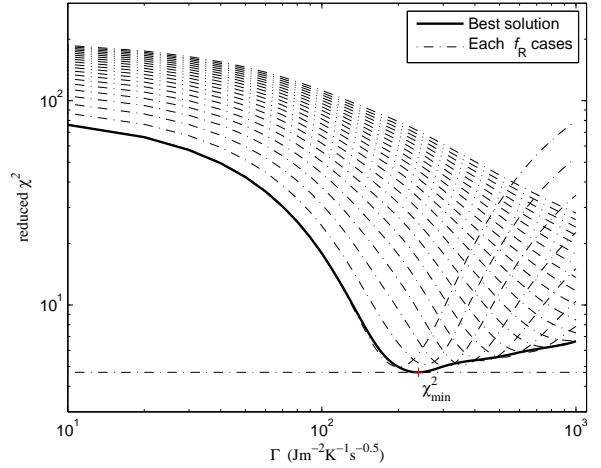}
  \centering
  \caption{$\Gamma\sim \chi^{2}$ profile fit to the observations.
  Each dashed curve represents a roughness fraction $f_{\rm R}$ in the
  range of $0.0\sim1.0$. The bold black line is a cubic spline interpolation
  curve for each lowest $\chi^{2}$ derived from each free parameter.
  }\label{chi2plot}
\end{figure}

To obtain the best-fitting solution of thermal inertia from Table
\ref{fitchi2}, the $\Gamma\sim \chi^{2}$ curves are plotted to
determine how $\chi^{2}$ globally changes with the free parameters of
thermal inertia, roughness fraction and effective diameter (see
Figure \ref{chi2plot}).

In Figure \ref{chi2plot}, the bold black curve represents the best
estimated solution in the fitting, which is a cubic spline
interpolation curve for each lowest $\chi^{2}$ calculated from each
thermal inertia and roughness fraction. From the curve, we find
$\chi_{\rm min}^{2}$ (referred to the minimum $\chi^{2}$), occurs in the
case of $\Gamma=240\rm~Jm^{-2}s^{-0.5}K^{-1}$, $f_{\rm R}=0.04$, and
$D_{\rm eff}=510\rm~m$. In addition, we note that $\chi_{\rm min}^{2}$
decreases rapidly when $\Gamma$ is less than
$240\rm~Jm^{-2}s^{-0.5}K^{-1}$, but increases rather slowly when
$\Gamma$ becomes larger than $240\rm~Jm^{-2}s^{-0.5}K^{-1}$. This
phenomenon results from the absence of observations at low phase
angle, and thus places limitations on the parameter space of
thermal inertia. However, the combination of observations at several
different phase angles actually weakens the abovementioned
disadvantage. Therefore, we can still achieve the best estimate
solution for thermal inertia and roughness for Bennu.

\begin{figure}
\includegraphics[scale=0.6]{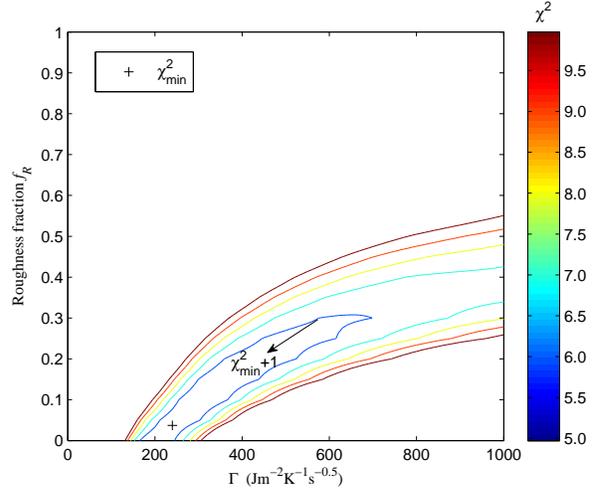}
  \centering
  \caption{$\chi^{2}$ ($\Gamma$, $f_{\rm R}$) contour according to Table \ref{fitchi2}.
  The color (from blue to red) means the increase of profile of $\chi^{2}$.
  The blue curve labeled by $\chi^{2}_{\rm min}+1$ is assumed to be 1$\sigma$ limit to the
  free fit parameters \citep{Emery2014,Bevington2003}.
  }\label{chi2contour}
\end{figure}

\begin{landscape}
\begin{table}
\vskip 50pt
\renewcommand\arraystretch{1.5}
\centering
\caption{ATPM fitting results to the observations.(The combined effective diameters $D_{\rm eff}$ are given in m)}
\label{fitchi2}
\begin{tabular}{@{}ccccccccccccccccccc@{}}
\hline
 Roughness &\multicolumn{12}{c}{Thermal inertia $\Gamma$ ($\rm~Jm^{-2}s^{-0.5}K^{-1}$)} \\
 fraction & \multicolumn{2}{c}{0} & \multicolumn{2}{c}{100}
          & \multicolumn{2}{c}{200} & \multicolumn{2}{c}{300}
          & \multicolumn{2}{c}{400} & \multicolumn{2}{c}{500}
          & \multicolumn{2}{c}{600} & \multicolumn{2}{c}{800}
          & \multicolumn{2}{c}{1000}\\
 $f_{\rm R}$ & $D_{\rm eff}$ & $\chi^{2}$ & $D_{\rm eff}$ & $\chi^{2}$
             & $D_{\rm eff}$ & $\chi^{2}$ & $D_{\rm eff}$ & $\chi^{2}$
             & $D_{\rm eff}$ & $\chi^{2}$ & $D_{\rm eff}$ & $\chi^{2}$
             & $D_{\rm eff}$ & $\chi^{2}$ & $D_{\rm eff}$ & $\chi^{2}$
             & $D_{\rm eff}$ & $\chi^{2}$ \\
\hline
 0.00 & \textbf{475.1} & \textbf{87.04} & \textbf{496.2} & \textbf{17.87} & \textbf{516.8} &  \textbf{4.97} & 535.2 &  9.49 & 550.6 & 20.00 & 563.3 & 31.85 & 573.7 & 43.55 & 589.9 & 63.77 & 601.1 & 79.70 \\
 0.05 & 457.1 & 98.04 & 478.0 & 24.37 & 497.6 &  6.48 & 514.8 &  5.85 & 529.3 & 11.51 & 541.1 & 19.06 & 550.8 & 26.96 & 565.9 & 41.08 & 576.3 & 52.50 \\
 0.10 & 441.0 & 108.07 & 461.6 & 31.04 & 480.2 &  9.45 & \textbf{496.5} &  \textbf{5.00} & 510.0 &  7.08 & 521.0 & 11.48 & 530.1 & 16.60 & 544.1 & 26.32 & 553.8 & 34.47 \\
 0.15 & 426.6 & 117.23 & 446.7 & 37.66 & 464.5 & 13.29 & 479.8 &  5.96 & \textbf{492.5} &  \textbf{5.36} & 502.8 &  7.40 & 511.3 & 10.46 & 524.3 & 16.90 & 533.4 & 22.64 \\
 0.20 & 413.5 & 125.59 & 433.2 & 44.11 & 450.2 & 17.63 & 464.6 &  8.09 & 476.5 &  5.45 & 486.2 &  5.69 & 494.2 &  7.17 & 506.3 & 11.15 & 514.8 & 15.05 \\
 0.25 & 401.5 & 133.25 & 420.9 & 50.30 & 437.1 & 22.20 & 450.8 & 10.97 & 462.0 &  6.74 & \textbf{471.1} &  \textbf{5.58} & \textbf{478.6} &  \textbf{5.86} & 489.9 &  7.93 & 497.9 & 10.42 \\
 0.30 & 390.6 & 140.26 & 409.5 & 56.21 & 425.1 & 26.87 & 438.1 & 14.30 & 448.7 &  8.82 & 457.3 &  6.58 & 464.3 &  5.90 & 474.9 &  6.47 & 482.4 &  7.85 \\
 0.35 & 380.5 & 146.72 & 399.1 & 61.81 & 414.0 & 31.52 & 426.4 & 17.89 & 436.4 & 11.43 & 444.5 &  8.31 & 451.2 &  6.87 & \textbf{461.2} &  \textbf{6.25} & 468.3 &  6.73 \\
 0.40 & 371.3 & 152.66 & 389.5 & 67.12 & 403.8 & 36.08 & 415.6 & 21.62 & 425.1 & 14.36 & 432.8 & 10.53 & 439.1 &  8.48 & 448.6 &  6.88 & \textbf{455.2} &  \textbf{6.64} \\
 0.45 & 362.7 & 158.16 & 380.5 & 72.14 & 394.3 & 40.53 & 405.6 & 25.39 & 414.7 & 17.48 & 422.0 & 13.07 & 427.9 & 10.52 & 436.9 &  8.13 & 443.2 &  7.28 \\
 0.50 & 354.8 & 163.24 & 372.2 & 76.88 & 385.5 & 44.82 & 396.3 & 29.14 & 405.0 & 20.70 & 411.9 & 15.81 & 417.6 & 12.85 & 426.1 &  9.78 & 432.1 &  8.43 \\
 0.55 & 347.4 & 167.97 & 364.5 & 81.36 & 377.3 & 48.95 & 387.7 & 32.83 & 395.9 & 23.95 & 402.6 & 18.66 & 408.0 & 15.35 & 416.1 & 11.72 & 421.8 &  9.94 \\
 0.60 & 340.5 & 172.36 & 357.2 & 85.58 & 369.6 & 52.92 & 379.6 & 36.44 & 387.5 & 27.18 & 393.9 & 21.56 & 399.0 & 17.95 & 406.8 & 13.84 & 412.3 & 11.70 \\
 0.65 & 334.0 & 176.46 & 350.5 & 89.57 & 362.5 & 56.72 & 372.0 & 39.94 & 379.7 & 30.38 & 385.7 & 24.47 & 390.7 & 20.60 & 398.1 & 16.08 & 403.4 & 13.62 \\
 0.70 & 328.0 & 180.29 & 344.1 & 93.35 & 355.7 & 60.35 & 365.0 & 43.32 & 372.3 & 33.50 & 378.2 & 27.35 & 382.9 & 23.27 & 390.0 & 18.38 & 395.0 & 15.65 \\
 0.75 & 322.4 & 183.87 & 338.2 & 96.92 & 349.4 & 63.82 & 358.4 & 46.59 & 365.4 & 36.55 & 371.0 & 30.18 & 375.6 & 25.91 & 382.4 & 20.71 & 387.3 & 17.73 \\
 0.80 & 317.1 & 187.24 & 332.6 & 100.30 & 343.5 & 67.13 & 352.2 & 49.74 & 359.0 & 39.51 & 364.4 & 32.95 & 368.7 & 28.52 & 375.3 & 23.04 & 380.0 & 19.84 \\
 0.85 & 312.1 & 190.40 & 327.3 & 103.50 & 338.0 & 70.30 & 346.3 & 52.77 & 352.9 & 42.37 & 358.1 & 35.65 & 362.3 & 31.07 & 368.6 & 25.34 & 373.1 & 21.95 \\
 0.90 & 307.4 & 193.38 & 322.4 & 106.54 & 332.7 & 73.32 & 340.8 & 55.68 & 347.2 & 45.14 & 352.2 & 38.28 & 356.3 & 33.56 & 362.4 & 27.61 & 366.7 & 24.04 \\
 0.95 & 303.0 & 196.18 & 317.7 & 109.43 & 327.8 & 76.21 & 335.7 & 58.47 & 341.8 & 47.81 & 346.7 & 40.82 & 350.6 & 35.99 & 356.5 & 29.84 & 360.7 & 26.11 \\
 1.00 & 298.9 & 198.84 & 313.4 & 112.18 & 323.2 & 78.97 & 330.8 & 61.15 & 336.8 & 50.38 & 341.5 & 43.28 & 345.3 & 38.35 & 351.0 & 32.01 & 355.1 & 28.14 \\
\hline
\end{tabular}
\end{table}
\end{landscape}

Figure \ref{chi2contour} shows a contour of $\chi^{2}$ in the
2-dimensional parameter space ($\Gamma$, $f_{\rm R}$), in which
$\chi^{2}$ is represented by colour. The increase of $\chi^{2}$ is
shown by ColorBar from blue to red. The black '+' shows where
$\chi_{\rm min}^{2}$ occurs in the ($\Gamma$, $f_{\rm R}$) parameter
space. The blue curve corresponds to $\chi_{\rm min}^{2}+1$, where a
1$\sigma$ limit of the free fit parameters $\Gamma$ and $f_{\rm R}$
is assumed. We observe that the blue profile is closed in the
($\Gamma$, $f_{\rm R}$) space, indicating that we can provide a
constraint for thermal inertia and roughness fraction
simultaneously, within the 1$\sigma$ limit. However, when
considering a higher limit of free parameters above 1$\sigma$, the
degeneracy of thermal inertia and roughness fraction cannot be
removed as well as 1$\sigma$ level, indicating that thermal inertia
and roughness fraction may be simply separated at 1$\sigma$ level
based on the calculations given in Table \ref{fitchi2}. Therefore,
if 1$\sigma$ limit is reliable, we may safely conclude that the
roughness fraction is likely to be in the range of 0$\sim$0.3,
whereas the thermal inertia is possibly in the range of
$180\sim680\rm~Jm^{-2}s^{-0.5}K^{-1}$. Our result coincides with
earlier investigations of \citet{Muller2012} and \citet{Emery2014}.
Table \ref{outcome} summarizes the derived results of the properties
of Bennu.

\begin{table}
 \centering
 \renewcommand\arraystretch{1.8}
 \caption{Derived properties of Bennu from ATPM.}
 \label{outcome}
 \begin{tabular}{@{}lc@{}}
 \hline
 Property                             & Result \\
 \hline
 Thermal inertia $\Gamma$
 ($\rm~Jm^{-2}s^{-0.5}K^{-1}$)        & $240^{+440}_{-60}$    \\
 Roughness fraction $f_{\rm R}$       & $0.04^{+0.26}_{-0.04}$   \\
 Effective diameter $D_{\rm eff}$~(m) & $510^{+6}_{-40}$   \\
 Geometric albedo $p_{v}$             & $0.047^{+0.0083}_{-0.0011}$ \\
 \hline
\end{tabular}
\end{table}

In the subsequent section, we will now employ these derived
parameters to evaluate the surface thermal environment of Bennu at
its aphelion and perihelion, respectively.

\subsection{Temperature Distribution}
Since Bennu is a spacecraft target asteroid, the surface temperature
is very essential to be estimated as a reference for the mission input.
Now the surface nature of Bennu's thermal inertia and geometric
albedo are precisely derived from the thermal modelling process,
therefore we can easily obtain the surface temperature distribution of
Bennu at any epoch from the thermophysical model.

\begin{figure*}
\includegraphics[scale=0.75]{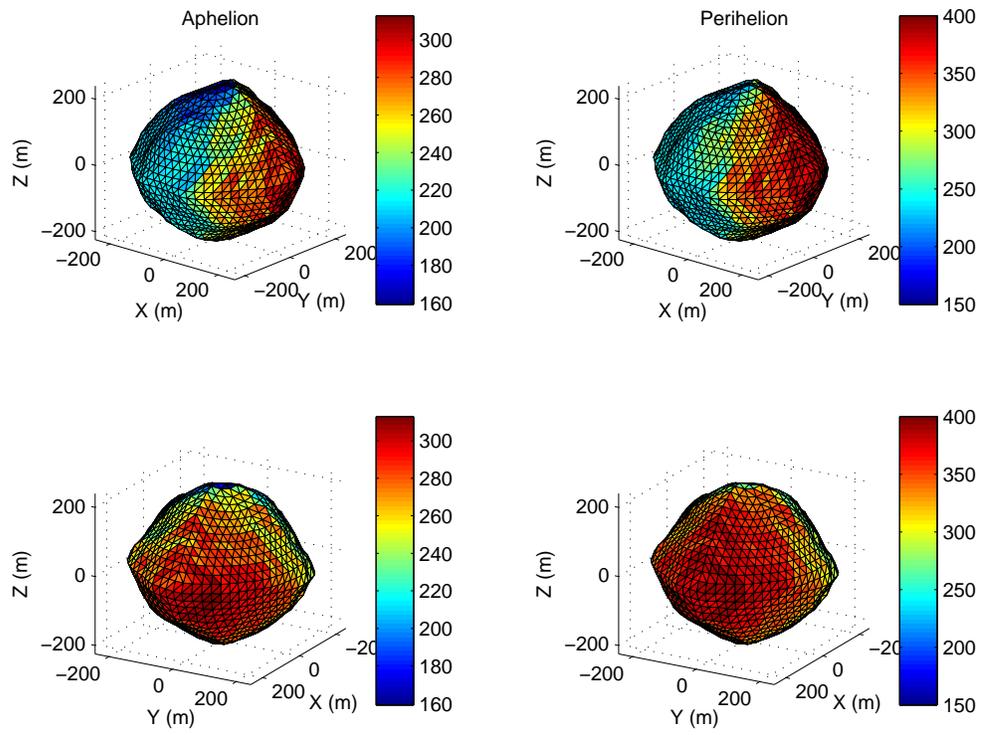}
  \centering
  \caption{Global surface temperature distribution simulated by ATPM based on
  the derived thermal inertia $\Gamma=240\rm~Jm^{-2}s^{-0.5}K^{-1}$ at
  the aphelion and perihelion, respectively. The colorbar indicates the range
  of temperature, where red for high temperature and blue for low temperature.
  }\label{tglobal}
\end{figure*}

\begin{figure*}
\includegraphics[scale=0.95]{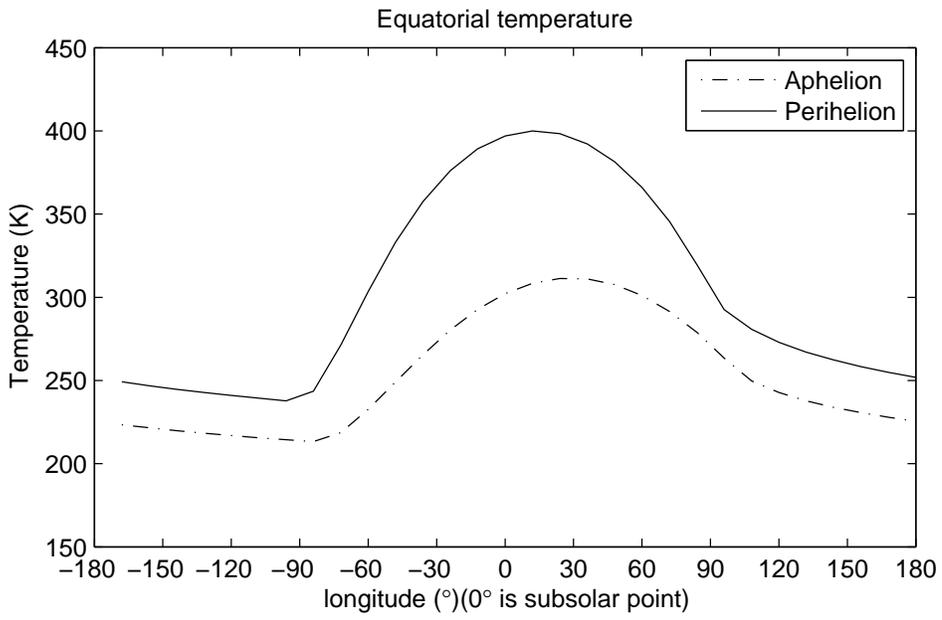}
  \centering
  \caption{Equatorial temperature distribution of Bennu at aphelion and
  perihelion, respectively.
  }\label{tequator}
\end{figure*}

Figure \ref{tglobal} shows the global surface temperature
distribution of Bennu at its aphelion (left panel) and perihelion
(right panel), respectively. In the Figure, the utilized-coordinate
can be named as "the frame system", where the coordinate origin
represents the asteroid center, the z-axis coincides with the
positive spin axis, and the x-axis follows the rules so that the Sun
always locates within the x-z-plane. The profile of temperature in
Figure \ref{tglobal} is shown by the colorbar index - the red region
represents the facets are sunlit, whereas the blue facets are
related to relatively low temperatures. As shown in Figure
\ref{tglobal}, the surface temperatures of its aphelion and
perihelion are roughly in the range of $160$ - $300\rm~K$ and $150$
- $400\rm~K$, respectively.

Figure \ref{tequator} shows the equatorial temperature distributions
of Bennu at its aphelion and perihelion, respectively. The maximum
temperature does not appear at the sub-solar point, but delays
$\sim$ $28^\circ$, and the minimum temperature occurs just a little
after the local sunrise, trailing $\sim$ $12^\circ$. This postponed
effect between absorption and emission is actually brought about by
non-zero thermal inertia and the finite rotation speed of the
asteroid. On the other hand, Figure \ref{tequator} shows that the
equatorial temperature of Bennu alters from $220$ to $400\rm~K$ over
an entire orbital period.

\subsection{Regolith}
As mentioned previously, Bennu has been chosen as the target asteroid
of the OSIRIS-REx sample return mission; thus we show great interest
in the surface features of the asteroid, whether a regolith layer exists
on its surface. Generally, thermal inertia is a good indicator to infer
the presence or absence of loose material on the asteroid's surface.
As is well known, fine dust has a very low thermal inertia of
$\sim$ $30\rm~Jm^{-2}s^{-0.5}K^{-1}$, and lunar regolith corresponds
to a relatively low value about $50\rm~Jm^{-2}s^{-0.5}K^{-1}$.
In comparison, the soil of a sandy regolith like Eros may have a value
of $100-200\rm~Jm^{-2}s^{-0.5}K^{-1}$, but coarse sand (e.g., Itokawa's
Muses-Sea Regio) is likely to possess a relatively higher thermal inertia
profile $\sim$ $400\rm~Jm^{-2}s^{-0.5}K^{-1}$. However, bare rock has an
extremely high thermal inertia, greater than $2500\rm~Jm^{-2}s^{-0.5}K^{-1}$
\citep{Delbo2007}. In summary, the given information suggests that a lower
thermal inertia for the asteroid may be related to a regolith layer. Since
the derived thermal inertia $\Gamma=240^{+440}_{-60}\rm~Jm^{-2}s^{-0.5}K^{-1}$
of Bennu is larger than that of Eros but lower than that Itokawa, we may
infer that regolith may exist on the surface of Bennu.

Although thermal inertia is associated with the surface properties,
a question arises -- what does the profile of thermal inertia tell us
about the surface properties? Generally, we are interested in the investigation
of grain size and regolith thickness of the asteroid, because these play
an important role in understanding the primitive substance on/underneath the
regolith layer; they further provide good engineering parameters
for the sample return mission. Theoretically, the so-called 'skin depth':
\begin{equation}
l_{s}=\sqrt{\frac{\kappa}{\rho c\omega}}=\frac{\Gamma}{\rho c\sqrt{\omega}},
\end{equation}
is usually presented to characterize the maximum grain size of
regolith. Thus, if the $\kappa$, $\rho$, $c$ and $\omega$ of Bennu
are known, in general \textbf{$l_{s}$ can be estimated to reveal the
maximum grain size of regolith on the surface of Bennu.} Recently,
the bulk density of Bennu has been updated to
$1.26\pm0.07\rm~g~cm^{-3}$ \citep{Chesley2014}. However, the
regolith density for Bennu is unknown, so the bulk density may work
as a reference because the regolith density is generally no larger
than the bulk density. Thus we adopt the bulk density as an
approximation for the average density of the surface regolith when
estimating the skin depth $l_{s}$. The rotation period is about
4.2976 h according to \citet{Nolan2013}. We may use the specific
heat capacity of CM or CI carbonaceous chondrites to approximate the
specific heat capacity of the surface regolith, where
$c\approx500\rm~Jkg^{-1}K^{-1}$ \citep{Opeil2012}. In this way, we
can estimate the skin depth of Bennu to be about 1.9 cm, which
suggests that the grain size may be less than cm-scale. Of course,
the estimation of grain size from skin depth is rough; thus it would
be appreciated if another way could be found to estimate the grain
size of Bennu's surface regolith.

According to the definition of thermal inertia:
\begin{equation}
\Gamma=\sqrt{\rho c\kappa}=\sqrt{\phi\rho_{\rm p}c\kappa}~,
\label{krcg}
\end{equation}
where $\rho$ is mean density of the surface, $\rho_{\rm p}$ is the grain's
density, $\phi$ (1-porosity) represents packing fraction of surface, $c$ is
the specific heat capacity and $\kappa$ is the thermal conductivity.
Thus the thermal conductivity $\kappa$ is directly related to the
thermal inertia $\Gamma$ and packing fraction $\phi$.

\citet{Gundlach2013} builds a thermal conductivity model and provides
a formula to theoretically estimate thermal conductivity $\kappa$ from
average radii of grains $r$, packing fraction $\phi$ and temperature $T$:
\begin{eqnarray}
\kappa(r,T,\phi)=\kappa_{\rm solid}[\frac{9\pi}{4}\frac{1-\mu^{2}}{E}\frac{\gamma(T)}{r}]^{1/3}
\cdot[f_{1}exp(f_{2}\phi)]\cdot\chi
\nonumber\\
+8\sigma\epsilon T^{3}e_{1}\frac{1-\phi}{\phi}r ~.
\label{krtphi}
\end{eqnarray}
The details of eq. (\ref{krtphi}) are described in \citep{Gundlach2013}.
Thus, the relationship between thermal inertia $\Gamma$ and grain size $r$
can be obtained by comparing the $\kappa$ derived from $\Gamma$ to the
$\kappa$ estimated from Formula (\ref{krtphi}). Hence, we can estimate
the surface grain radius $r$ with the thermal inertia derived from
above-mentioned thermophysical modelling process.

\begin{figure}
\includegraphics[scale=0.58]{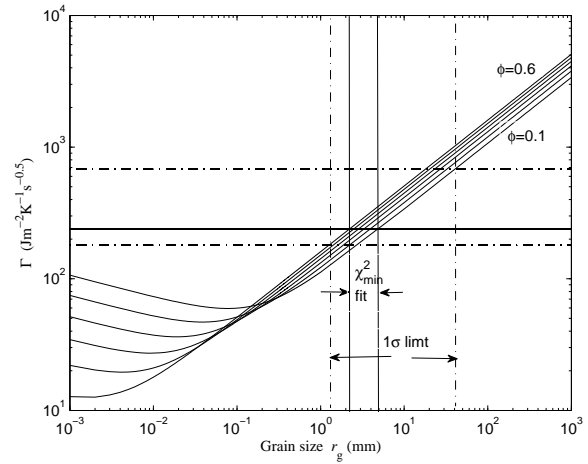}
  \centering
  \caption{$r\sim\Gamma$ curve of Bennu based the method of \citet{Gundlach2013},
  giving the possible range of grain radius $r$ between $2\sim5$~mm for the best
  fit thermal inertia $240\rm~Jm^{-2}s^{-0.5}K^{-1}$. While the 1$\sigma$ range of
  $\Gamma$ is considered, the possible grain radius $r$ may be between $1.3\sim31$~mm .}
  \label{grainsize}
\end{figure}

Figure \ref{grainsize} shows that the $r\sim\Gamma$ curve is plotted with
the parameters listed in Table \ref{krpa} in combination of Equation \ref{krcg}
and \ref{krtphi}. We assume an average surface temperature $T=300$~K based on
Figure \ref{tequator} in the computation. The other parameters are adopted
from \citet{Gundlach2013}. Since there are still many assumptions and uncertainties
in the parameters of Table \ref{krpa}, the uncertainties of (\ref{krtphi}) are rather
difficult to determine. Thus we just choose the suggested parameters to estimate the
most likely results for Bennu. Using the best-fit value of thermal inertia
$\Gamma=240\rm~Jm^{-2}s^{-0.5}K^{-1}$, we estimate the grain radius is
likely to be in the range $2\sim5$~mm. Furthermore, considering a 1$\sigma$ range
of thermal inertia, we may estimate the grain radius is possibly in the range
between $1.3\sim31$~mm. According to this evaluation of grain radius, we infer that
boulders or rocks may be few on the surface of Bennu, implying that the
Touch-And-Go Sample Acquisition Mechanism (TAGSAM) designed by the OSIRIS-REx team
will be available to be implemented successfully.

\begin{table}
 \centering
 \renewcommand\arraystretch{1.2}
 \caption{Physical parameters in eq. \ref{krtphi} \citep{Gundlach2013}.}
 \label{krpa}
 \begin{tabular}{@{}cc@{}}
 \hline
 Property & Value  \\
 \hline
 $\kappa_{\rm solid}$ & $1.19+2.1\times10^{-3}T$~[$\rm Wm^{-1}K^{-1}$]  \\
 $\mu$      &     0.25    \\
 $E$        &     $7.8\times10^{10}$~[Pa]  \\
 $\gamma(T)$ &  $6.67\times10^{-5}T$~[$\rm Jm^{-2}$]   \\
 $f_{1}$    &     $0.0518\pm0.0345$                   \\
 $f_{2}$    &     $5.26\pm0.94$                       \\
 $\chi$     &     $0.41\pm0.02$                       \\
 $e_{1}$    &     $1.34\pm0.01$                       \\
 $\epsilon$ &      1                                  \\
 $\rho_{\rm p}$  &  3110~[$\rm kgm^{-3}$]  \\
 $c$             &  560~[$Jkg^{-1}K^{-1}$] \\
 $T$        &      300~K \\
 \hline
 \multicolumn{2}{l}{$\kappa_{\rm solid}$: thermal conductivity of the solid material} \\
 \multicolumn{2}{l}{$\mu$: Poisson's ratio}\\
 \multicolumn{2}{l}{$E$: Young's modulus}\\
 \multicolumn{2}{l}{$\gamma(T)$: specific surface energy} \\
 \multicolumn{2}{l}{$\epsilon$: emissivity of the material} \\
 \multicolumn{2}{l}{$\rho_{\rm p}$: density of the solid material} \\
 \multicolumn{2}{l}{$c$: heat capacity of the solid material} \\
 \multicolumn{2}{l}{$T$: surface temperature} \\
\end{tabular}
\end{table}

\section{Discussion and Conclusion}
In this work, using the thermophysical model incorporating the 3D
radar-derived shape model \citep{Nolan2013}, and mid-infrared data
of Spitzer-PUI, Spitzer-IRAC, Herschel/PACS and ESO VLT/VISIR
observations \citep{Muller2012,Emery2014}, we have investigated the
thermophysical nature of Bennu. We show that the thermal inertia
$\Gamma=240^{+440}_{-60}\rm~Jm^{-2}s^{-0.5}K^{-1}$, the roughness
fraction $f_{\rm R}=0.04^{+0.26}_{-0.04}$, the effective diameter
$D_{\rm eff}=510^{+6}_{-40}\rm~m$, and the geometric albedo $p_{\rm
v}=0.047^{+0.0083}_{-0.0011}$. The effective diameter acquired
herein is a little larger than the former results
\citep{Muller2012,Emery2014} and that of the radar-derived shape
model \citep{Nolan2013}, but the value of geometric albedo we
derived is much closer to those of \citet{Muller2012} and
\citet{Emery2014}, because a greater absolute visual magnitude is
used in the fitting process. The best solution of thermal inertia is
even lower than that of \citet{Emery2014}, whereas the low roughness
results are consistent with each other. The slight deviation of
Bennu's thermal inertia between this work and \citet{Emery2014} may
arise from three aspects: first of all, we adopt a different
thermophysical modelling procedure \citep{Rozitis2011,Yu2014}, where
the beaming is modelled by solving the energy conservation equation
in consideration of multiple scattering of sunlight and self-heating
of thermal emission; then, the model-derived fluxes are rotationally
averaged in this work, whereas \citet{Emery2014} constrained the
rotational phase directly from Spitzer data; last, the thermal
inertia and roughness fraction are simultaneously determined in the
2D parameter space, thus leading to a larger possible range of
thermal inertia than that of \citet{Emery2014}. In a word, the
physical parameters for Bennu we derived are essentially supportive
of those of \citet{Emery2014}.

\begin{figure}
\includegraphics[scale=0.6]{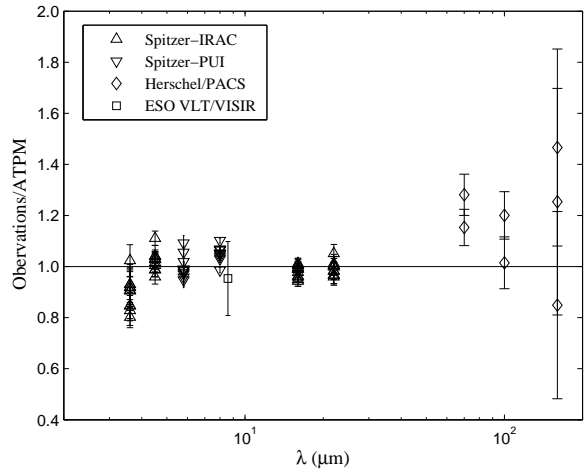}
\centering
\caption{The observation/ATPM ratios as a function of wavelength
  for $f_{\rm R}=0.04$, $\Gamma=240\rm~Jm^{-2}s^{-0.5}K^{-1}$,
  and $D_{\rm eff}=510\rm~m$.
  }\label{rspectra}
\end{figure}

\begin{figure}
\includegraphics[scale=0.6]{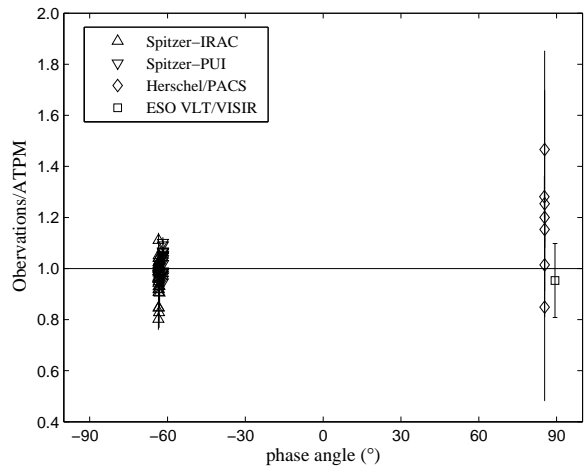}
\centering
\caption{The observation/ATPM ratios as a function of phase angle
  for $f_{\rm R}=0.04$, $\Gamma=240\rm~Jm^{-2}s^{-0.5}K^{-1}$,
  and $D_{\rm eff}=510\rm~m$.
  }\label{rphase}
\end{figure}

We employ the ratio of 'observation/model' \citep{Muller2005,Muller2011,Muller2012}
to examine how the theoretical model results match the observations at various
phase angles and wavelengths (see Figure \ref{rspectra} and \ref{rphase}), for
the reliability of our fitting process and derived outcomes may be verified from
these comparisons.

In Figure \ref{rspectra}, the observation/ATPM ratios are shown at each
observational wavelength for $f_{\rm R}=0.04$, $\Gamma=240\rm~Jm^{-2}s^{-0.5}K^{-1}$,
and $D_{\rm eff}=510\rm~m$. The ratios are distributed nearly symmetrically
around 1.0, despite several ratios at low-wavelength $3.6\rm~\mu m$ and long-wavelength
about $100\rm~\mu m$ move relatively farther from unity. From Figure \ref{rphase},
we find that the deviations of low-wavelength and long-wavelength come from Spitzer-IRAC
and Herschel/PACS, respectively. As the Wien peak of the thermal emission of NEAs
approximately arises at $10\rm~\mu m$, the deviation at those
wavelengths far away from the Wien peak is inevitable, indicating that Herschel observations
at long-wavelength \citep{Muller2012} are not very sensitive to the averaged surface
temperature of Bennu, and thus have insignificant influence on the derived thermal inertia.
However, the Herschel data can still be helpful to improve the determination of the effective diameter.

On the other hand, although the Spitzer-IRAC and Herschel/PACS observations appear to
deviate away from unity, the degeneracy of thermal inertia and roughness are actually
removed via our modelling procedure through the combination fitting to all observations
(see Table \ref{obs}) at discrepant phase angles. Therefore, we successfully provide a
constraint for thermal inertia and roughness fraction simultaneously based on the 1$\sigma$
limit. This processing method differs from previous models, which usually determine
thermal inertia with several empirical roughnesses. Hence, we safely come to
the conclusion that the outcomes of thermal inertia and roughness are reliably derived
from our modelling process.


In conclusion, the average thermal inertia of Bennu of approximately
$\Gamma=240^{+440}_{-60}\rm~Jm^{-2}s^{-0.5}K^{-1}$, provides direct
evidence that the asteroid's surface properties are an intermediate
case between Eros and Itokawa, implying the possible existence of
regolith on its surface. As described above, a very small roughness
fraction for the asteroid is derived simultaneously with thermal
inertia in the modelling procedure, therefore we infer that Bennu's
surface would be fairly smooth. Additionally, based on the
best-estimate thermal inertia and roughness, fine-grained regolith
would be likely to exist and cover a large area of the surface of
Bennu. Finally, the estimated grain size ranging from $1.3$ to
$31$~mm is indicative that Bennu would be an excellent target
asteroid for the forthcoming OSIRIS-REx sample return mission.

\section*{Acknowledgments}
The authors thank the anonymous referee for his/her constructive
comments that greatly helped to improve the content of this
manuscript. This work is financially supported by National Natural Science
Foundation of China (Grants No. 11273068, 11473073, 11403105), the Strategic
Priority Research Program-The Emergence of Cosmological Structures
of the Chinese Academy of Sciences (Grant No. XDB09000000), the
innovative and interdisciplinary program by CAS (Grant No.
KJZD-EW-Z001), the Natural Science Foundation of Jiangsu Province
(Grant No. BK20141509), and the Foundation of Minor Planets of
Purple Mountain Observatory.

\label{lastpage}

\end{document}